\documentstyle[preprint,aps,psfig]{revtex}

\newcommand{\be}{\begin{eqnarray}}
\newcommand{\ee}{\end{eqnarray}}
\newcommand{\bc}{\begin{center}}
\newcommand{\ec}{\end{center}}

\begin{document}
\draft
\tighten

\title { La p\^eche dans les trous noires}
\author{ A. Brotas\footnote{E-mail:brotas@fisica.ist.utl.pt}   }
\address{ Departamento de F\'{\i}sica, Instituto Superior T\'ecnico, \\
Av Rovisco Pais 1096.  Lisboa Codex, Portugal}

\date{\today}

\maketitle

\begin{abstract}

	The problem of a thread whose extremity is beyond the Schwarzschild 
horizon is discussed. 

	We present a new coordinate system allowing to write the 
Schwarzschild metric in all regions.

	The system of coordinates introduced appears to be particulary 
indicated to study the passage of extended bodies throuth the Schwarzschild 
horizon, as it is the case of a thread that continues to attach a 
fisherman to a fish that entered a black hole. \\  

\begin{large} \bc Resum\'e \ec \end{large}

	On discute le probl\`eme d'un fil dont l'extr\'emit\'e d\'epasse 
l'horizon de Schwarzschild $r_h = 2m$ ;

\bc $ ds^2 = - dr^2 \left( 1 - \frac{2m}{r} \right)^{-1} - r^2 \left( 
d\theta^2 + \sin^2\theta d\phi^2 \right) + \left( 1 - \frac{2m}{r} \right) 
c^2 dt^2 $ \ec 

	Le changement de coordonn\'ees:

\bc $ r = r_0 - cK \left( \overline{\tau} - \frac{\overline{x}}{B_0} \right) 
~~;~~ B_0 = cK \left( 1 - \frac{2m}{r_0} + K^2 \right)^{-\frac{1}{2}} 
~~;~~ \left( r_0 \ge 2m ~,~ 0 < K^2 < \infty \right) $ \ec 

\bc  \[
t = \frac{\overline{x}}{cK} - \frac{1}{cK} \int^{r}_{r_0} 
\left( 1 - \frac{2m}{r} + K^2 \right)^{\frac{1}{2}} \left( 1 - \frac{2m}{r}
 \right)^{-1} dr \] \ec
la coordonn\'ee $\overline{x}$ \'etant la longuer d'un fil inextensible 
et $\overline{\tau}$ le temps marqu\'e par des horloges li\'es aux 
points du fil, permet d'\'ecrire la m\'etrique de Schwarzschild sous 
la forme:

\bc $ ds^2 = \left[ \frac{c^2}{B_0} + \frac{1}{K^2} - \frac{2m}{K^2r} - 
\frac{2c}{KB_0} \left( 1 - \frac{2m}{r} + K^2 \right)^{\frac{1}{2}} \right] 
d\overline{x}^2 -  r^2 \left( d\theta^2 + \sin^2\theta d\phi^2 \right) 
 + 2\left[ -\frac{c^2}{B_0} + \frac{c}{K} \left( 1 - \frac{2m}{r} + K^2 
 \right)^{\frac{1}{2}} \right] d\overline{x} d\overline{\tau} + c^2dt^2 
 $ \ec
qui peut \^etre utiliz\'e dans toute la region $r > \frac{2m}{1+K^2}$ .

\end{abstract}
\newpage

	On sait que l'{\it horizon de Schwarzschild} est une {\it 
patologie} li\'ee au choix des coordonn\'ees habituellement utilis\'ees 
dans l'\'ecriture de la m\'etrique de Schwarzschild:

\be ds^2 = - dr^2 \left( 1 - \frac{2m}{r} \right)^{-1} - r^2 \left( 
d\theta^2 + \sin^2\theta d\phi^2 \right) + \left( 1 - \frac{2m}{r} \right) 
c^2 dt^2 \ee 

	Eddington, Kruskal, Novikov et d'autres ont present\'e des syst\`emes 
de coordonn\'ees pour lequeles il n'y a pas en $r = r_h = 2m$ aucune sorte 
de singularit\'e et il est possible d'\'etudier le mouvement des 
particules (et des photons) qui traversent $r_h = 2m$ et tombent en $r=0$. \\

	La distance {\it statique} (\`a $t = ct.$) entre un point 
$r_0 > r_h$ et $r_h = 2m$ donn\'ee par:

\be d(r_h,r_0) = - \int_{r_0}^{r_h} \left( 1 - \frac{2m}{r} \right)^{
-\frac{1}{2}} dr \ee
\'etant finie, il se pose d'embl\'ee le probl\`eme de savoir ce qui arrive 
\`a un fil suspendu en $r_0$ et dont l'extr\'emit\'e p\'en\`etre en $r_h$ 
(il suffit pour cela d'avoir une longuer $L$ sup\'erieur \`a $d(r_h,r_0)$. 
Nous pouvons imaginer, par example, que le fil sort avec une vitesse 
constante d'un treuil plac\'e en $r_0$. \\

	Une particule partie de $r_0$ en chute libre ou en mouvement 
forc\'e ne peut attendre $r_h$ avant $t = \infty $. De plus, une fois 
arriv\'ee en $r_h$, la particule tombe, in\'evitablemente, en $r=0$. 
Un point mat\'eriel ayant atteint $r_h$ ne peut donc retourner \`a $r_0$. \\ 

	Comment, dans ces conditions, imaginer un fil dont l'extr\'emit\'e 
d\'epasse $r_h$ ? Nous pouvons imaginer que le treuil qui a laiss\'e 
sortir le fil renverse sa marche. Comment concilier cette possibilit\'e 
(toujour imaginable) d'inversion de la marche du treuil avec 
l'impossibilit\'e de r\'ecuperation du fil ? Comment imaginer le 
simple cas d'un fil avec une extremit\'e fixe en $r_0$ et l'autre 
tombant in\'evitablement en $r=0$ apr\`es avoir d\'epass\'e $r_h$ ? \\

	Si nous travaillons avec un fil inextensible (rigide-ind\'eformable 
\`a une dimension aux sens de Born) nous nous trouvons, sans aucune 
doute, devant des situations paradoxales. Mais ces paradoxes sont 
semblables \`a bien d'autres qui surgissent en Relativit\'e Restreite 
et en Relativit\'e G\'en\'erale quand nous utilizons le corps 
rigide-ind\'eformable de Born (simple notion g\'eometrique) comme 
mod\`ele des objects physiques \'etendus que nous imaginons en 
mouvement dans nos exp\'eriences mentales. \\

	Les paradoxes disparaisent quand nous considerons des corps 
d\'eformables sumis \`a des lois \'elastiques. \\ 

	Les lois \'elastiques relativistes \cite{um}, \cite{dois}, 
\cite{tres}, \cite{quatro}, y compris celle qui correspond au cas limite 
des corps dans lequels les ondes de choc se d\'eplacente avec la 
vitesse $c$ (que j'appellerai les v\'eritables corps rigides \cite{cinco}):

\be p = \frac{\rho^0_0 c^2}{2} \left( \frac{1}{s^2} - 1 \right) ~~~;~~~ 
s = \frac{L}{L_0} ~~~~ (\rho^0_0 ~ densit\acute{e} ~ propre ~ iniciale) \ee
ont cette propri\'et\'e curieuse; la pression $p$ tends vers une valeur 
finie quand $L$ tends vers l'infini. On peut donc \'etirer de fa\c{c}on 
illimit\'ee un fil, m\^eme si \`a la limite il est rigide. Le fait qu'une
extremit\'e du fil tombe en $r=0$ est ainsi parfaitement conciliable 
avec la possibilit\'e pour l'autre d'\^etre fixe en $r_0$, ou m\^eme 
d'\^etre receuillie par un treuil. \\

	Un p\^echer qui laisse un poisson entrer dans un trou noir 
pourra, par la suite, rembobiner le moulinet pour un temp illimit\'e, 
sans r\'eussir \`a l'en faire sortir ni l'emp\^echer de tomber \`a 
$r=0$, tout en le tenant toujours au but de sa ligne. \\

	Etudions, pour le moment, le mouvement d'un fil inextensible 
sorti avec une vitesse constante d'un treuil plac\'e en $r_0$. \\

	Soit $\Delta \bar{x} = cK\Delta t$ la quantit\'e (longueur 
propre) du fil qui sort du treuil dans l'intervale $\Delta t$ (dans 
la direction de $r=0$). On trouve (apr\`es des calculs que nous 
expliciterons par la suite) que la quantit\'e du fils en mouvement 
comprise entre $r_1 > r_h$ et $r_0$ (que nous pouvons appeler 
distance {\it dynamique}) est donn\'ee par:

\be d_k (r_1,r_0) = \int_{r_1}^{r_0} 
\left( 1 - \frac{2m}{r} + K^2 \right)^{\frac{1}{2}}
\left( 1 - \frac{2m}{r} \right)^{-1} dr \ee

	Nous constatons que $d_K (r_h,r_0)$ a une valeur infinie 
(m\^eme pour des $K$ tr\`es petits). Le treuil peut donc travailler 
pendant un temps infini sans que "du point de vue de $t$" l'extremit\'e 
du fil arrive \`a $r_h$. (ce r\'esultat est bien d'accord avec 
l'impossibilit\'e pour un point mat\'eriel d'atteindre $r_h$ dans 
un temps $t$ fini, ce qui ne signifie pourtant pas que le temps 
propre pour un voyage de $r_0$ \`a $r_h$ soit n\'ecessairemente 
infini). \\

	Cherchons l'equation $\bar{x} = \bar{x}(r,t)$ que d\'ecrit 
le mouvement du fil. La coordonn\'ee $\bar{x}$ sera la longuer 
propre du fil compt\'ee \`a partir du point $\bar{x} = 0$ qui 
passe en $r_0$ \`a l'instant $t=0$. \\

	La condition d'inextensibilit\'e impose que les quantit\'es 
de fil $\Delta \bar{x}$ qui passent en $r_0$ et $r_1$ dans le 
m\^eme intervale $\Delta t$ soient les m\^emes. Nous avons ainsi:

\be \left( \frac{\partial \bar{x}}{\partial t} \right)_{r_1}  = 
\left( \frac{\partial \bar{x}}{\partial t} \right)_{r_0} = cK ~~~;
~~~ (K = ct.) \ee

	Les observateurs imobiles en $r$ voient passer les points 
du fil avec la vitesse:

\be v = \left( \frac{\partial \bar{x}}{\partial t} \right)_{r} 
\left( 1 - \frac{v^2}{c^2} \right)^{\frac{1}{2}} 
\left( 1 - \frac{2m}{r} \right)^{-\frac{1}{2}} \ee 

	En faisant les calculs nous obtenons:

\be v = cK \left( 1 - \frac{2m}{r} + K^2 \right)^{-\frac{1}{2}} ~~~; 
~~~ (0 < v < c ) ~~ \Rightarrow ~~ (0 < K^2 < \infty ) \ee

	En calculant la "distance dynamique" $d_K (r,r_0)$ \`a partir de:

\be d_K (r,r_0) = \int_r^{r_0} \left( 1 - \frac{v^2}{c^2} \right)^{-\frac{1}{2}} 
\left( 1 - \frac{2m}{r} \right)^{-\frac{1}{2}} dr \ee 
et en utilizant (5) et (7) nous obtenons (4). \\ 

	La quantit\'e de fil qui est pass\'ee en $r$ \`a l'instant 
$t$ est celle qui est sorti du treuil dans l'intervale $[0,t]$ 
moins celle qui est entre $r$ et $r_0$ (\`a $t=ct.$). Nous pouvons donc 
\'ecrire: 
\be \bar{x} (r,t) = cKt - d_K (r,r_0) \ee 

	Pour calculer le temps propre $d \tau$ necessaire pour qu'un point 
du fil ($\bar{x}=ct.$) parcoure $dr$ nous utilizons $cd\tau = ds$ 
et la relation:

\be 0 = cKdt + \left( 1 - \frac{2m}{r} + K^2 \right)^{\frac{1}{2}}
\left( 1 - \frac{2m}{r} \right)^{-1} dr \ee
obtenue de (9) pour le cas $\bar{x}=ct.$. Ces expressions nous donnent:

\be dr = - c K d \tau \ee
r\'esultat particuli\`erement simple. Le temps propre necessaire pour 
qu'un point du fil aille de $r_0$ \`a $r_1$ est donc:

\be r_1 = r_0 - cK\Delta \tau = r_0 - cK (\tau_1 - \tau_0) \ee

	Admettons que les points du fil qui sort du treuil plac\'e en 
$r_0$ sont munis d'horloges (qui marquent le temps propre $\tau$) et 
que ces horloges sont r\'egl\'es au passage de $r_0$ conformement 
\`a la formule:

\be \tau = t \left( 1 - \frac{v^2}{c^2} \right)^{-\frac{1}{2}} 
\left( 1 - \frac{2m}{r_0} \right)^{\frac{1}{2}} \ee
(Ce processus de reglage fait que les horloges du fil soient r\'egl\'es entre eux 
au passage en $r_0$). En utilisant (7) et (9) nous obtenons:

\be \tau = \frac{\bar{x}}{B_0} ~~~~;~~~~ B_0 = cK \left( 1 - \frac{2m}{r_0} + K^2 
\right)^{-\frac{1}{2}} \ee

Representons par $\bar{\tau}$ le temps marqu\'e par les horloges du fil 
r\'egl\'es par le processus indiqu\'e. En utilisant les formules (12), (14), 
(9) et (8) nous pouvons \'ecrire:

\be \left\{ \begin{array}{l} 
r = r_0 - ck \left( \bar{\tau} - \frac{\bar{x}}{B_0} \right) \\  \\
t = \frac{\bar{x}}{cK} - \frac{1}{cK} \int_{r_0}^{r} \left( 1 - \frac{2m}{r} + K^2 
\right)^{\frac{1}{2}} \left( 1 - \frac{2m}{r} \right)^{-1} dr \end{array} \right. \ee

Pour un $K$ donn\'e $(\bar{x},\bar{\tau})$ est un syst\`eme de coordonn\'ees qui peut 
\`etre utilis\'e dans le r\'egion $r > 2m$. En calculant les $g_{\alpha \beta}$ 
respectives nous obtenons:

\be \begin{array}{l}
g_{\bar{\tau} \bar{\tau}} = c^2 ~~~~;~~~~ 
g_{\bar{x} \bar{\tau}} =  -\frac{c^2}{B_0} + \frac{c}{K} \left( 1 - \frac{2m}{r} + K^2 
 \right)^{\frac{1}{2}} \\ \\ 
g_{\bar{x} \bar{x}} =  \frac{c^2}{B_0} + \frac{1}{K^2} - \frac{2m}{K^2r} - 
\frac{2c}{KB_0} \left( 1 - \frac{2m}{r} + K^2 \right)^{\frac{1}{2}} \end{array} \ee 

Ces coeficients sont definis et n'ont pas de singularit\'es pour $r >
2m \left( 1+K^2 \right)^{-1}$. Le calcul direct du tenseur de Ricci donne 
$R_{\alpha \beta}=0$. 
Le syst\`eme $(\bar{x},\bar{\tau})$ peut donc \^etre utilis\'e en alternative \`a 
$(r,t)$ non seulement dans la region $r>2m$, mais dans toute la r\'egion 
 $r > \frac{2m}{1+K^2}$. En choisissant un $K$ suffisament grand ont peut approcher
cette r\'egion du centre $r=0$ autant qu'on le veut. 

	Le premier r\'esultat (16) \'etait pr\'evisible \'etait donn\'e que 
$\bar{\tau}$ est un temps propre. En utilisant les suivants et en faisant les 
calculs nous obtenons: 

\be \gamma_{\bar{x} \bar{x}} = g_{\bar{x} \bar{x}} - g_{\bar{x} \bar{x}}^2 ~~
g_{\bar{\tau} \bar{\tau}}^{-1} = -1 \ee
ce qui traduit la condition que nous avons impos\'ee d\`es le debut, \`a savoir, 
que $\bar{x}$ soit la longueur d'un fil inextensible.

	En faisant en (14) $r_0 = 2m$, ce qui correspond \`a mettre le treuil \`a 
$r_h = 2m$ (ce qui est impossible), ou \`a le mettre ailleurs mais \`a r\'egler 
les horloges de fa\c{c}on telle qu'ils soient accord\'ees au passage en 
$r_h$, nous obtenons $B_0 = c$. Dans ce cas, les formules (15) et (16) 
prenent la forme plus simple;

\be \left\{ \begin{array}{l} 
r = r_0 + K \bar{x} - cK \bar{\tau}  \\  \\
t = \frac{\bar{x}}{cK} - \frac{1}{cK} \int_{r_0}^{r_0 + K \bar{x} - Kc \bar{\tau}}
 \left( 1 - \frac{2m}{r} + K^2 
\right)^{\frac{1}{2}} \left( 1 - \frac{2m}{r} \right)^{-1} dr \end{array} \right. \ee

et 

\be \begin{array}{l}
g_{\bar{\tau} \bar{\tau}} = c^2 ~~~~;~~~~ 
g_{\bar{x} \bar{\tau}} =  - c + \frac{c}{K} \left( 1 - \frac{2m}{r} + K^2 
 \right)^{\frac{1}{2}} \\ \\ 
g_{\bar{x} \bar{x}} =  \frac{1}{K^2}  \left( 1 - \frac{2m}{r} + K^2 \right)
 - \frac{2}{K} \left( 1 - \frac{2m}{r} + K^2  \right)^{\frac{1}{2}} 
\end{array} \ee 

	En $r=r_h = 2m$, ces coefficients prennent les valeurs 	
$g_{\bar{\tau} \bar{\tau}} = c$ ; $g_{\bar{x} \bar{\tau}} = 0$ et 
$g_{\bar{x} \bar{x}} = -1$, ce qui montre que les coordonn\'ees 
 $(\bar{x},\bar{\tau})$ sont localement les coordonn\'ees d'un 
syst\`eme euclidien associ\'e au fil. Elles sont donc particuli\`erement 
propices \`a l'\'etude des objects \'etendus qui entrent 
dans les trous noires. \footnote[1]
{	Madame Choquet-Bruhat, membre de l'Academie des Sciences de Paris, 
m'a donn\'e l'honneur de pr\'esenter ce texte dans une s\'eance de 
l'Academie, en 1982. Il n'a pas \'et\'e publi\'e dans les Comptes Rendues 
parce que un "referee" a consid\'er\'e ce syst\`eme de coordon\'ees sans 
int\'er\^et.

	 Dans la suite, il a \'et\'e utilis\'e pour \'ecrire l'\'equation 
du passage d'un fil "rigide" par l'horizon de Schwarzschild et \'etudier 
ses solutions. Je ne connais pas d'autres \'etudes concernant l'entr\'e 
des corps \'etandus dans les trous noires.}


\begin{references}


\bibitem{um} Mc Crea, Sci. Proc. R. Dublin Soc. (N.S.), 26 (1952); \\
               Hogart and Mc Crea, Proc. Cambr.Phil. Soc. 48 (1952)
\bibitem{dois} A. Brotas, "Sur le probl\`eme du disque tournant", \\
                  C.R. Acad. Sc. Paris, t. 267 A, 57 (1968)
\bibitem{tres} A. Brotas, Th\`ese Paris 1969 (N enregistrement C.N.R.S. A.O. 3081)
\bibitem{quatro} A. Brotas and J.C. Fernandes, "A lei de Hooke relativista", \\
                 T\'ecnica 461, Lisboa (1980)
\bibitem{cinco} A. Brotas, "{\it Rigide} et {\it ind\'eformable} sont-ils des synonimes?", \\
                 L. Nuovo Cimento  I 1, 217 (1969);
\bibitem{seis} A. Brotas and J. C. Fernades, The relativistic elasticity of rigid 
             bodies, arXiv:physics/0307019 v2 19 Set 2003



\end{references}
\end{document}